\documentclass[twocolumn,10pt,superscriptaddress,amsmath,amssymb,aps,prb]{revtex4-2}
\usepackage{amsmath}
\usepackage{graphicx}
\usepackage{dcolumn}
\usepackage{bm}
\usepackage{mathtools}
\usepackage{xcolor}
\usepackage{hyperref}
\hypersetup{colorlinks=true,breaklinks,linkcolor=blue,urlcolor=blue,citecolor=blue}

\begin{document}

\title{Spin current generation controlled by the N\'{e}el state in a compensated ferrimagnet}

\author{Xin Theng Lee}
\affiliation{Institute for Solid State Physics, University of Tokyo, Kashiwa, 277-8581, Japan}

\author{Takahiro Misawa}
\affiliation{Institute for Solid State Physics, University of Tokyo, Kashiwa, 277-8581, Japan}

\author{Mamoru Matsuo}
\affiliation{Kavli Institute for Theoretical Sciences, University of Chinese Academy of Sciences, Beijing, China}
\affiliation{CAS Center for Excellence in Topological Quantum Computation, University of Chinese Academy of Sciences, Beijing, China}
\affiliation{Advanced Science Research Center, Japan Atomic Energy Agency, Tokai, Japan}
\affiliation{RIKEN Center for Emergent Matter Science (CEMS), Wako, Saitama, Japan}

\author{Takeo Kato}
\affiliation{Institute for Solid State Physics, University of Tokyo, Kashiwa, 277-8581, Japan}

\date{\today}

\begin{abstract}
Compensated ferrimagnets, which break sublattice and time-reversal symmetries in the ground state, exhibit an isotropic ferromagnet-like spin splitting despite a vanishing net magnetization, in contrast to altermagnets with momentum-dependent spin splitting.
We investigate how isotropic spin splitting manifests in spin transport by analyzing the spin Seebeck effect and spin pumping in a junction between a compensated ferrimagnet and a normal metal.
We show that compensated ferrimagnets generate a sizable spin Seebeck signal,
with a sign that can be reversed
by switching between the two N\'{e}el states.
Furthermore, we demonstrate that spin pumping exhibits a N\'{e}el-state-dependent resonance splitting, which is absent in conventional antiferromagnets.
These results identify spin pumping as a natural readout mechanism for compensated ferrimagnets and establish them as promising magnetization-free building blocks for spintronic memory devices.
\end{abstract}

\maketitle

\section{Introduction}
Antiferromagnetic spintronics  offers high-density integration and ultrafast dynamics for information technology owing to the absence of stray magnetic fields~\cite{Jungwirth2016,Baltz2018,Jungwirth2018}.
However, a fundamental bottleneck persists. The very property that makes antiferromagnets attractive, namely zero net magnetization, renders efficient electrical writing and reading inherently challenging.
These operations typically rely on electrical signals that are weak compared to the strong responses of ferromagnets.
To overcome this trade-off, current efforts seek a material platform that combines the absence of stray magnetic fields characteristic of antiferromagnets with the readability of ferromagnets.
While the recent discovery of altermagnets~\cite{Noda_PCCS2016,Naka2019,Ahn_PRB2019,Hayami2020,Smejkal2022,Smejkal_PRX2022b,Mazin2022,Yuan2023,Cui2023,Hodt2024} has challenged the conventional notion that zero magnetization implies spin degeneracy, their momentum-dependent (e.g., $d$-wave) spin splitting imposes geometric constraints on charge and spin transport.
Here, we propose a complementary solution: compensated ferrimagnets (CFs).
Unlike altermagnets, CFs exploit sublattice inequivalence to generate \textit{isotropic} ($s$-wave) spin splitting~\cite{Leuken_PRL1995, Akai_PRL2006, Kawamura_PRL2024, Liu_PRL2025}, thereby mimicking ferromagnetic transport properties while maintaining zero net magnetization.

Compensated ferrimagnets provide a platform that combines two key advantages: the absence of stray magnetic fields characteristic of antiferromagnets and a macroscopic spin-splitting energy scale comparable to that of ferromagnets, thereby enabling strong spin responses.
This concept has recently moved beyond purely theoretical proposals, as advances in materials synthesis, including organic compounds~\cite{Kawamura_PRL2024} and two-dimensional van der Waals heterostructures~\cite{Liu_PRL2025}, have enabled controlled sublattice asymmetry at the atomic scale.
Despite this progress, the connection to device-relevant phenomena, in particular spin transport across junctions, remains largely unexplored. Since junctions constitute the fundamental building blocks of spintronic devices, clarifying how isotropic spin splitting in compensated ferrimagnets translates into measurable transport signals is essential for realizing efficient stray-field-free memory devices.

In this paper, we theoretically demonstrate that junctions between a compensated ferrimagnet and a normal metal can generate sizable spin signals, thereby addressing a central readout challenge in antiferromagnetic spintronics.
We analyze the spin Seebeck effect (SSE)~\cite{Uchida2008,Jaworski2010,Slachter2010,Uchida2011,Jaworski2012,Adachi2013} and spin pumping (SP)~\cite{Tserkovnyak2002a,Tserkovnyak2002b,Tserkovnyak2005,Saitoh2006} using the nonequilibrium Green's function method.
We show that sublattice asymmetry arising from unequal easy-axis anisotropies induces isotropic magnon spin splitting, leading to a SSE signal whose sign can be reversed via transitions between two N\'{e}el states.
We also find that spin pumping exhibits a N\'{e}el-state-dependent shift in the resonance frequency, a feature absent in conventional antiferromagnets.
These results establish compensated-ferrimagnet--normal-metal junctions as a platform for nonequilibrium spin transport and readout and indicate their potential for stray-field-free spintronic devices with reliable signal generation.

\begin{figure}[tb]
    \centering
    \includegraphics[width=1.0\linewidth]{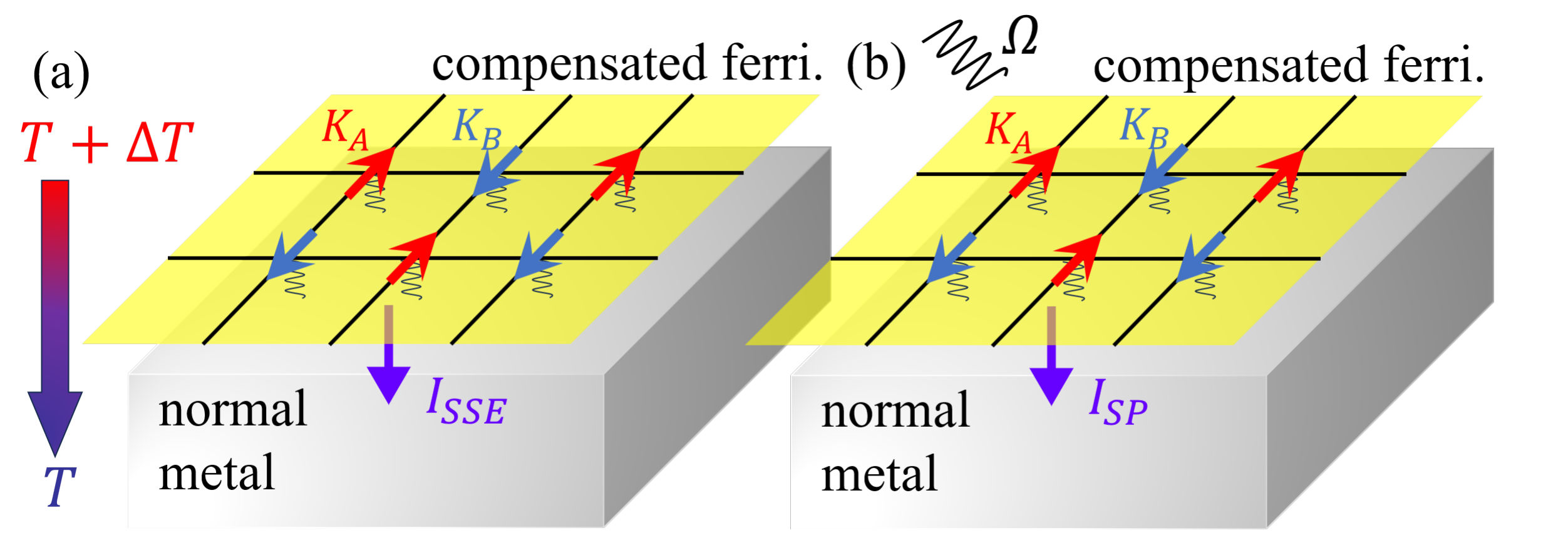}
    \caption{Schematic setups for spin-current generation in a compensated ferrimagnet (CF)--normal metal (NM) junction. Red and blue arrows represent spins on sublattices A and B, with different easy-axis anisotropies $K_{\rm A}$ and $K_{\rm B}$. (a) Spin Seebeck effect: a temperature difference $\Delta T$ between the CF at $T+\Delta T$ and the NM at $T$ drives a spin current $I_{\rm SSE}$ across the interface. (b) Spin pumping: a microwave field of angular frequency $\Omega$ drives coherent spin precession in the CF, injecting a spin current $I_{\rm SP}$ into the NM at zero temperature difference.}
\label{fig:schem}
\end{figure}

\section{Model and Method}
\subsection{Compensated ferrimagnets}
\label{subsec:CFerri}

We consider a two-dimensional square-lattice model of a collinear antiferromagnet with different easy-axis anisotropies, $K_{\rm A}$ and $K_{\rm B}$, on the two sublattices (see Fig.~\ref{fig:schem}).
The Hamiltonian is given by
\begin{align}
    \label{eq:compensatedferrimodel}
    H_{\rm CF} = \sum_{\langle i, j \rangle} J \bm S_i \cdot \bm S_j - K_{\rm A}\sum_{i \in {\rm A}} (S_i^z)^2 - K_{\rm B}\sum_{j \in {\rm B}} (S_j^z)^2
\end{align}
where $\bm S_i = (S_i^x,S_i^y,S_i^z)$ is the spin operator with magnitude $S_0$ and $\sum_{\langle i , j \rangle}$ indicates the sum over nearest-neighbor bonds connecting sites $i$ and $j$.
We assume a collinear N\'{e}el order with $\langle S_i^z \rangle = m S_0$ on sublattice A and $\langle S_j^z \rangle = -m S_0$ on sublattice B, where $m = \pm 1$ indicates two degenerate N\'{e}el configurations.
To analyze the excitation spectrum for the $m=+1$ configuration,
we employ the standard spin-wave approximation based on the Holstein-Primakoff transformation~\cite{HolsteinPrimakoff1940}:
\begin{align}
S_{\nu \bm{R}}^z &=
\begin{cases}
S_0 - a_{\bm{R}}^{\dagger}a_{\bm{R}}, & (\nu = \rm A), \\
- S_0 + b_{\bm{R}}^{\dagger}b_{\bm{R}}, & (\nu = \rm B),
\end{cases} \label{eq:spinwaveapprox1} \\
S_{\nu \bm{R}}^+ & \approx
\begin{cases}
\sqrt{2S_0}a_{\bm{R}}, & (\nu = \rm A), \\
\sqrt{2S_0}b_{\bm{R}}^\dagger, & (\nu = \rm B),
\end{cases}
\label{eq:spinwaveapprox2} \\
S_{\nu \bm{R}}^- & = (S_{\nu \bm{R}}^+)^\dagger , \label{eq:spinwaveapprox3}
\end{align}
where $S_{\nu \bm{R}}^z$ and $S_{\nu \bm{R}}^\pm = S_{\nu \bm{R}}^x \pm i S_{\nu \bm{R}}^y$ are spin operators at the site ${\bm R}$ on the sublattice $\nu$, and $a_{\bm R}$ ($b_{\bm R}$) is an annihilation operator of magnons on sublattice A (B).
Then, the Hamiltonian is rewritten in a quadratic form with respect to the magnon creation and annihilation operators:
\begin{align}
    \label{modelequation_main}
    H_{\text{CF}} &= \sum_{\bm{k} \in \Lambda}
    (a_{\bm k}^\dagger \ b_{-{\bm k}})
    \left( \begin{array}{cc} A & C_{\bm k} \\
    C_{\bm k} & B \end{array} \right) \left( \begin{array}{c} a_{\bm k} \\ b_{-{\bm k}}^\dagger \end{array} \right),
\end{align}
where $a_{\bm k}$ and $b_{{\bm k}}$ are the Fourier transforms of $a_{{\bm R}}$ and $b_{\bm R}$, respectively, $A = 4JS_0 + 2K_{\rm A} S_0$, $B = 4JS_0 + 2K_{\rm B} S_0$, $C_{\bm k} = JS_0\sum_{\bm \delta} e^{i\bm k\cdot\bm\delta}$
($\bm \delta$ denotes the nearest-neighbor vectors), and $\Lambda$ indicates the first Brillouin zone.
Using a Bogoliubov transformation, $a_{\bm k} = u_{\bm k} \alpha_{\bm k} - v_{\bm k}\beta_{-\bm k}^\dagger$ and
$b_{-\bm k}^\dagger = -v_{\bm k} \alpha_{\bm k} + u_{\bm k}\beta_{-\bm k}^\dagger$, where $\alpha_{\bm k}$ and $\beta_{\bm k}$ are bosonic operators~\cite{Anderson1952,Auerbach1994}, we obtain the diagonalized Hamiltonian
\begin{align}
    & H_{\text{CF}} = \sum_{\bm{k} \in \Lambda}  \bigg(\hbar \omega_{+,\bm k} \alpha^{\dagger}_{{\bm k}} \alpha_{{\bm k}} + \hbar (-\omega_{-,\bm k})  \beta^{\dagger}_{{-\bm k}} \beta_{{-\bm k}} \bigg),
    \label{eq:diagonalized_main} \\
    &
    \hbar \omega_{{\pm},{\bm k}}
= S_0 \Delta K \pm 2 S_0 \sqrt{(2J + \bar{K})^2 -4 J^2 \gamma_{\bm k}^2} ,
\end{align}
where $\Delta K = K_{\rm A} - K_{\rm B}$, $\bar{K}=(K_{\rm A}+K_{\rm B})/2$, and $\gamma_{\bm k} = \sum_{\bm\delta} e^{i\bm k\cdot\bm\delta}/4 = (\cos k_x  a + \cos k_y a)/2$. We choose the labeling such that $\omega_{-,\bm k}<0<\omega_{+,\bm k}$. The magnon frequencies $\omega_{{\pm},{\bm k}}'$ for the spin configuration of $m=-1$ can be calculated in a similar way and are given by $\omega_{{\pm},{\bm k}}' = - \omega_{{\mp},{\bm k}}$ from the symmetry under the transformation $(m,{\bm S}_{\rm A},{\bm S}_{\rm B}) = (-m,{\bm S}_{\rm B},{\bm S}_{\rm A})$.

We present magnon band dispersions for $K_{\rm A} = 0.08J$ and $K_{\rm B} = 0.04J$ in Fig.~\ref{fig:BandStructure} (see Appendix~\ref{app:banddiagram} for the definition of the momentum path shown in the inset).
The magnon modes exhibit isotropic spin splitting with an amplitude of $2|\Delta K|S_0$.
This is a key feature of compensated ferrimagnets~\cite{Liu_PRL2025} and leads to characteristic spin transport properties, as discussed below.

\begin{figure}[tb]
\centering
\includegraphics[width=1.0\linewidth]{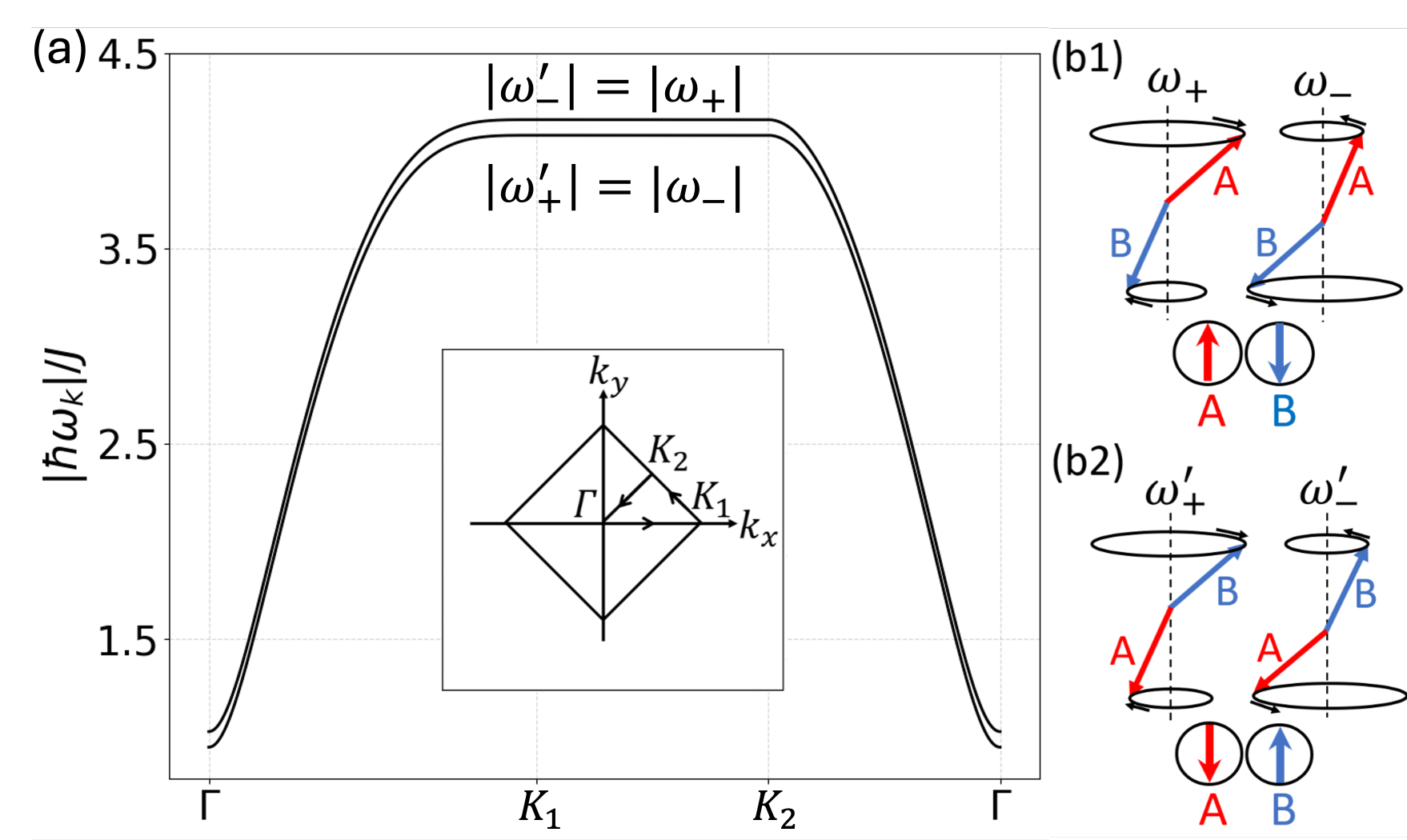}
\caption{(a) Magnon dispersion of a compensated ferrimagnet for $K_{\rm A} = 0.08J$ and $K_{\rm B} = 0.04J$, exhibiting uniform isotropic spin splitting along the momentum paths shown in the inset. (b1),(b2) Schematic illustration of the spin precession of the two magnon modes at the $\Gamma$ point (${\bm k}={\bm 0}$).}
\label{fig:BandStructure}
\end{figure}

It is instructive to consider the solutions of the Landau-Lifshitz-Gilbert (LLG) equation (see Appendix~\ref{app:magdyn}), which describe the magnon modes at the $\Gamma$ point (${\bm k}={\bm 0}$).
For the $m=+1$ configuration, the precession amplitude on sublattice A is larger than that on sublattice B for the $+$ mode, whereas it is smaller for the $-$ mode, as schematically shown in Fig.~\ref{fig:BandStructure} (b1).
For the spin configuration of $m=-1$, the amplitude relationship between the sublattices A and B is reversed, as shown in Fig.~\ref{fig:BandStructure} (b2).
In terms of the magnon approximation, the precession amplitudes obtained in the LLG equation correspond to the amplitudes of the wavefunctions, $u_{\bm k}$ and $v_{\bm k}$.
It is an essential feature of compensated ferrimagnets that the two magnon excitations have opposite net spins in the $z$ direction
due to the difference in precession amplitudes.

\subsection{Hamiltonian of the CF--NM junction}
The Hamiltonian of a junction between CF and a normal metal (NM) is given by
\begin{align}
H = H_{\text{CF}} + H_{\text{NM}} + H_{\text{int}}.
\label{Htot}
\end{align}
The first term $H_{\text{CF}}$ describes the compensated ferrimagnet and is given in Eq.~(\ref{eq:compensatedferrimodel}),
while $H_{\text{NM}}$ and $H_{\text{int}}$ describe the normal metal and the interfacial exchange coupling, respectively, and are given by
\begin{align}
H_{\text{NM}} &= \sum_{{\bm k}\sigma} \xi_{\bm k} c_{{\bm k}\sigma}^\dagger c_{{\bm k}\sigma}, \\
H_{\text{int}} &= \sum_{\bm{R}} J_{{\bm R}}\bm{S}_{\bm{R}} \cdot \bm{s}_{\bm r}, \label{dotprod}
\end{align}
where $\xi_{\bm k}$ represents the energy dispersion in the normal metal, $c_{{\bm k}\sigma}$ is an annihilation operator for conduction electrons with wavenumber ${\bm k}$ and spin $\sigma$, and
${\bm r}$ denotes the position in the normal metal neighboring the site ${\bm R}$ in the compensated ferrimagnet. The spin operator $\bm{s}_{\bm r}$ of the normal metal is defined as
\begin{align}
\bm s_{\bm r}
&= \frac{1}{2}c_{\bm r}^{\dagger}\bm\sigma c_{\bm r}
= \frac{1}{2N_{\rm NM}}
\sum_{\bm k,\bm q}\sum_{\sigma,\sigma'}
e^{i\bm q\cdot\bm r}
c_{\bm k\sigma}^{\dagger}
\bm\sigma_{\sigma\sigma'}
c_{\bm k+\bm q,\sigma'}.
\end{align}
where ${\bm \sigma}=(\sigma_x,\sigma_y,\sigma_z)$ are the Pauli matrices.

\subsection{Spin current}

The spin current flowing into the NM is given by
\begin{align}
I_S &= \frac{i}{2}\sum_{\nu,\bm{R}} (J_{\nu{\bm R}} S_{\nu\bm{R}}^{+} s_{{\bm r}}^- - {\rm h.c.}).
\end{align}
Using the Keldysh formalism~\cite{Kamenev2011,Keldysh1965},
the average spin current can be calculated to second order in the interfacial exchange coupling $H_{\rm int}$~\cite{kato19}.
We then perform a disorder average over the interfacial exchange coupling, assuming
\begin{align}
\left\langle J_{\nu{\bm R}} J_{\nu^{\prime}{\bm R}'}^{\ast}\right\rangle_{\rm imp} = |J_{\rm CF}|^2 \delta_{\bm{R},\bm{R}^{\prime}} \delta_{\nu,\nu^{\prime}} ,
\label{randave}
\end{align}
where $\langle \cdots \rangle_{\rm imp}$ denotes the disorder average over interfacial configurations, and $J_{\rm CF}$ represents the interfacial exchange coupling strength.
Finally, the average spin current is given by
\begin{align}
    \langle I_S \rangle
    &= \frac{\cal A}{N_{\rm CF}} \sum_{\nu,{\bm k}} \int_{-\infty}^{\infty} \frac{d(\hbar \omega)}{2 \pi} \, \hbar \omega   \nonumber \\
    &\times \text{Im}\, G_{\nu\nu}^R(\bm k,\omega)  \big(f_{\nu}^{\rm CF}(\hbar \omega) - f^{\rm NM}(\hbar \omega)\big),
    \label{eq:generalspincurrent_main}
\end{align}
where ${\cal A}=\pi N_{\rm int}|J_{\rm CF}|^2 N(0)^2/2$, $N_{\rm int}$ is the number of exchange bonds, $N(0)$ is the density of states of the normal metal at the Fermi energy, $N_{\rm CF}$ is the number of unit cells in CF, $G_{\nu\nu}^R(\bm k,\omega)$ is the retarded spin correlation function for the CF (see Appendix~\ref{appex:GreenFunction}), and $f_{\nu}^{\rm CF}(\hbar \omega)$ and $f^{\rm NM}(\hbar \omega)$ are the nonequilibrium distribution functions of spin excitations in CF and NM (for details, see Appendix~\ref{appex:derivationofSpinCurrent}).

\section{Results}
\subsection{Spin Seebeck effect}
We first consider the spin Seebeck effect, which generates a spin current across the interface between CF and a normal metal in the presence of a temperature difference $\Delta T$ (see Fig.~\ref{fig:schem}(a)).
Assuming that CF and NM are in thermal equilibrium at temperatures $T+\Delta T$ and $T$, respectively, the difference between the distribution functions is approximated to linear order in $\Delta T$ as
\begin{align}
\label{BoseLinearTemperature}
f_{\nu}^{\text{CF}}(\hbar \omega) - f^{\text{NM}}(\hbar \omega) \approx   \frac{\beta^2 \hbar \omega}{4\sinh^2(\beta \hbar \omega /2)} k_{\rm B} \Delta T ,
\end{align}
where $\beta = (k_{\rm B}T)^{-1}$.
The average spin current is given by~(Appendix~\ref{appex:derivationofSpinCurrent})
\begin{align}
\label{eq:SSEspincurrent}
I_{\rm SSE}^{\rm CF}
&= \frac{1}{N_{\rm CF}} \sum_{\bm{k}\in \Lambda}
\frac{S_0 \mathcal{A} k_{\rm B} \Delta T}{\sqrt{1 - \zeta_{\bm k}^2}} \notag \\
& \times \bigg(\frac{(\beta\hbar  \omega_{-,\bm k}/2)^2}{\sinh^2(\beta \hbar \omega_{-,\bm k}/2)} - \frac{(\beta\hbar \omega_{+,\bm k}/2)^2}{ \sinh^2(\beta\hbar \omega_{+,\bm k}/2)}\bigg),
\end{align}
where $\zeta_{\bm k}=2C_{\bm k}/(A+B)$.
This expression reflects the thermal energy imbalance that drives the spin Seebeck current across the CF--NM interface.
Eq.~(\ref{eq:SSEspincurrent}) shows that when the two sublattices A and B are equivalent (i.e., $K_{\rm A}=K_{\rm B}$), the spin current vanishes since $|\omega_{+,\bm k}|=|\omega_{-,\bm k}|$.
This result is consistent with the vanishing of the spin Seebeck effect in conventional antiferromagnets with equal interfacial exchange couplings on the two sublattices at zero external magnetic field~\cite{Masuda2024}.

\begin{figure}[tb]
\centering
\includegraphics[width=0.9\linewidth]{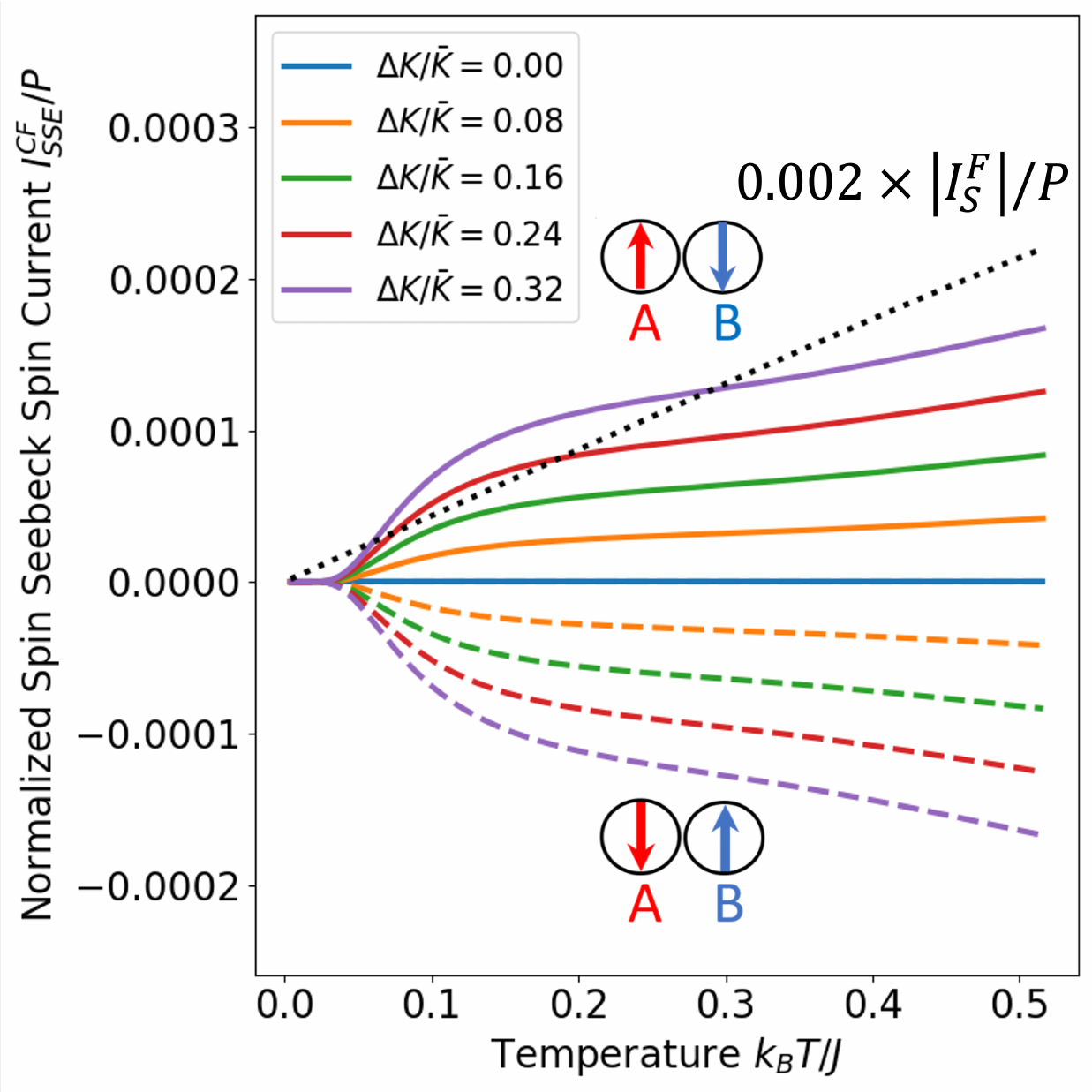}
\caption{Temperature dependence of the spin current in the CF--NM junction normalized by $P=4S_0 {\cal A}k_{\rm B}\Delta T$.
The parameters are chosen as $K_{\rm A}=\bar{K}+\Delta K/2$ and $K_{\rm B}=\bar{K}-\Delta K/2$, and the solid lines correspond to $\Delta K/\bar{K} = 0, 0.08, 0.16, 0.24$, and 0.32 for $\bar{K} = 0.005J$.
The dashed lines correspond to the case where the magnetizations on the A and B sublattices are flipped for finite $\Delta K$.
These curves are obtained by exchanging the anisotropy parameters between sublattices A and B.
We use $S_0=1$. The dotted line shows \(0.002\times|I_{\rm S}^{\rm F}|/P\) for \(J_{\rm FI}/J=0.3\), assuming the same interfacial coefficient \(\mathcal A\) for the CF and FI junctions.}
\label{fig:SSEcurrent(I)}
\end{figure}

We present the generated spin current as a function of temperature in Fig.~\ref{fig:SSEcurrent(I)}.
The spin current is approximately proportional to $\Delta K = K_{\rm A}-K_{\rm B}$ and varies monotonically with temperature.
The spin current is reversed when the magnetizations on both sublattices are flipped, because this operation exchanges the anisotropy parameters between the two sublattices (see the dashed lines in Fig.~\ref{fig:SSEcurrent(I)}).
Thus, the sensitivity of the spin current sign to the sublattice labeling provides an effective means to directly distinguish between the two N\'{e}el states.

This spin Seebeck effect can be understood as follows. The inequivalent energies of the two magnon modes lead to an imbalance in their thermal populations. For the $m=+1$ configuration, the $+$ mode lies at higher energy than the $-$ mode. As a result, the contribution of the $-$ mode to the interfacial spin transfer dominates, giving rise to a net $z$-polarized spin current flowing into the normal metal. For the opposite spin configuration ($m=-1$), the relative energies of the two modes are interchanged, resulting in a spin current flowing in the opposite direction.

For comparison, we consider the spin Seebeck effect in a junction between a ferromagnetic insulator (FI) and a normal metal (NM).
The spin current is calculated in the same manner as for the CF--NM junction, leading to
\begin{align}
\label{ferromagneticspinseebeck}
I_{\rm S}^{\rm F}
&= \frac{S_0 \mathcal A k_{\rm B} \Delta T}{N_{\rm FI}} \sum_{\bm{k}\in \Lambda}\frac{(\beta \lambda_{\bm{k}}/2)^2}{ \sinh^2(\beta \lambda_{\bm{k}}/2)} ,
\end{align}
where $\lambda_{\bm{k}} = D|\bm{k}|^2$ is the magnon dispersion in the FI~\cite{Kittel2005}, $D = J_{\text{FI}}S_0 a^2$ is the spin stiffness, $a$ is the lattice spacing, and $J_{\text{FI}}$ is the exchange coupling in the bulk FI.
For example, at $k_{\rm B}T/J=0.5$ and
$\Delta K/\bar K=0.32$, the CF spin Seebeck current is approximately $1.5\times10^{-3}$ of the FI current magnitude under the comparison conditions used in Fig.~\ref{fig:SSEcurrent(I)}.
We note that the total magnetization of the CF develops a finite value at finite temperatures.
However, its magnitude remains small; for example, it is approximately \(2.7\times10^{-4}\) per unit cell at \(k_{\rm B}T/J=0.85\) and \(\Delta K/\bar K=0.32\) (see Appendix~\ref{app:magnetization} and Fig.~\ref{fig:magnetization}).

\subsection{Spin pumping}

Next, we consider spin pumping induced by ferromagnetic resonance (FMR) driven by a microwave field in a junction between CF and NM coupled by an interfacial exchange interaction (see Fig.~\ref{fig:schem}(b)).
The total Hamiltonian is given by
\begin{align}
H = H_{\rm CF} + V(t) + H_{\rm NM} + H_{\rm int}.
\end{align}
The second term describes a circularly-polarized microwave field and is given by
\begin{align}\label{externalfield}
    V(t) &= -\frac{\hbar \gamma h_{ac}}{2}\sum_{\nu \bm{R}}\bigg(S_{\nu \bm{R}}^-e^{- i \Omega t} + {\rm h.c.}\bigg) \nonumber \\
    &\approx \mathcal{D} \bigg(a_{\bm{0}}^\dagger e^{- i \Omega t} + b_{\bm{0}}e^{- i \Omega t} + {\rm h.c.} \bigg)
\end{align}
where $h_\text{ac}$ and $\Omega$ ($>0$) are the amplitude and frequency of the microwave field, respectively, $\gamma$ is the gyromagnetic ratio, and $\mathcal{D} = -(\hbar \gamma h_{\text{ac}}/2) \sqrt{2S_0N_{\rm CF}}$.
A negative frequency ($\Omega<0$) corresponds to a microwave field with opposite circular polarization.
Then, the difference in the distribution functions is  determined by the uniform nonequilibrium magnon excitation and is given by
\begin{align}\label{eq:fluc-diss-SP}
& f^{\text{CF}}_{\nu}(\hbar \omega) -f^{\text{NM}}(\hbar \omega) \simeq \frac{\delta G^<_{\nu\nu}({\bm k},\omega)}{2i\,{\rm Im} \, [G_{\nu\nu}^R({\bm k},\omega)]} ,
\end{align}
where $\delta G^<_{\nu\nu}({\bm k},\omega)$ is a correction to the lesser component of the spin correlation function due to the microwave field.
To second order in ${\cal D}$, $\delta G^<_{\nu\nu}({\bm k},\omega)$ is calculated as
\begin{align}
\label{eq:pumpinglesserGF_main}
    & \delta G^<_{\nu\nu}(\bm{k},\omega) = - \frac{ i \pi{\cal D}^2}{S_0\hbar} \delta_{\bm k,\bm 0}\delta(\omega - \Omega)|\sum_\mu G^R_{\nu\mu}(\bm k,\omega)|^2 ,
\end{align}
Substituting Eqs.~(\ref{eq:fluc-diss-SP}) and (\ref{eq:pumpinglesserGF_main}) into Eq.~(\ref{eq:generalspincurrent_main}), the average spin current is calculated as
\begin{align}
\label{eq:spinpumpingspincurrent}
I_{\rm SP}(\Omega)
= -\frac{\mathcal A\mathcal D^2}{4S_0N_{\rm CF}}
\sum_{\nu}\hbar\Omega
|
\sum_{\mu}G_{\nu\mu}^{R}(\bm 0,\Omega)
|^2.
\end{align}
For the explicit forms of $G_{\nu\mu}^R(\bm{k},\Omega)$, see Appendix~\ref{appex:GreenFunction}.

\begin{figure}[tb]
    \centering
    \includegraphics[width=0.95\linewidth]{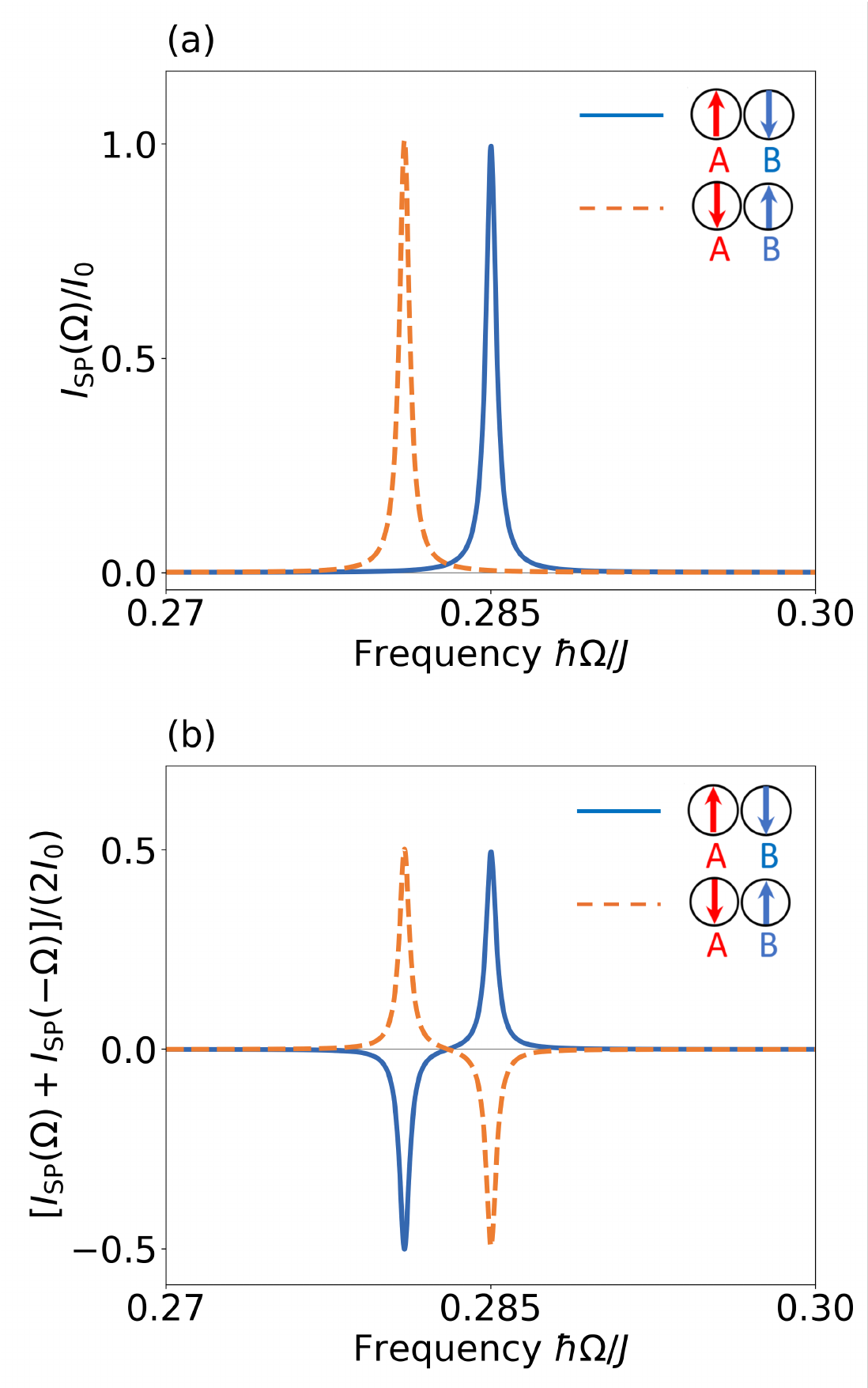}
    \caption{Normalized spin current generated by spin pumping as a function of frequency $\Omega$ for $K_{\rm A}=0.006J$, $K_{\rm B}=0.004J$, $S_0=1$, with the effective damping parameter $\tilde{\alpha}_{\rm G}$ set to $10^{-3}$ (see Appendix~\ref{appex:GreenFunction}). 
    (a) Spin current for a circularly polarized microwave field, $I_{\rm SP}(\Omega)/I_0$.  (b) Spin current for a linearly polarized microwave field, $(I_{\rm SP}(\Omega) + I_{\rm SP}(-\Omega))/2I_0$. 
    The circularly and linearly polarized fields are compared at equal time-averaged squared field amplitude.}
    \label{fig:SP}
\end{figure}

In Fig.~\ref{fig:SP}(a), the spin current generated by spin pumping in CF is shown as a function of frequency $\Omega$ for a circularly polarized microwave field ($\Omega>0$).
The spin current is normalized by $I_0 = -{\cal A}{\cal D}^2 S_0/(2N_{\rm CF}\tilde{\alpha}^2_{\rm G}\tilde{E})$, where $\tilde{E} = 2S_0 \sqrt{\bar{K}(4J+\bar{K})}$.
We observe a single FMR peak at $\Omega=\omega_{+,{\bm 0}}$ in the spin current, as shown by the solid line in Fig.~\ref{fig:SP}(a).
However, for the $m=-1$ configuration, the FMR frequency shifts to $\Omega=\omega'_{+,{\bm 0}}=|\omega_{-,{\bm 0}}|$, as shown by the dashed line in Fig.~\ref{fig:SP}(a).
This shift reflects the selective response of magnon excitations to the circular polarization of the microwave field (see also Fig.~\ref{fig:BandStructure}(b1), (b2)).
For linearly polarized microwave excitation, the spin current is given by the sum of the contributions for $\Omega$ and $-\Omega$ and exhibits two FMR peaks at $|\omega_{\pm,{\bm 0}}|$ with opposite signs, as shown in Fig.~\ref{fig:SP}(b).
By contrast, for conventional antiferromagnets with $K_{\rm A}=K_{\rm B}$ the spin current is not generated by a linearly polarized microwave field due to cancellation between the two magnon modes.
Thus, the two spin configurations in CF can be distinguished via the spin current induced by spin pumping.

\section{Discussion}

\subsection{Comparison with other magnetic ordering}

Our results identify compensated ferrimagnets as a distinct class of antiferromagnets that overcome key limitations of both altermagnets and conventional antiferromagnets in spintronic applications.
In particular, the \textit{isotropic} ($s$-wave) spin splitting in CFs eliminates the geometric constraints inherent to altermagnets with anisotropic spin splitting~\cite{Smejkal_PRX2022b}.
As a result, spin currents can be generated independently of the transport direction, enabling the use of polycrystalline samples and complex device geometries.
This flexibility substantially broadens the range of viable materials and device architectures for stray-field-free spintronics.
In Ref.~\cite{Hodt2024}, it is shown that for a homogeneous magnetic field, the spin current pumped from an altermagnet is identical to that from a conventional antiferromagnet. To obtain an altermagnet-specific signal, a spatially inhomogeneous field and sublattice-dependent interfacial exchange coupling are required. In CFs, by contrast, the isotropic spin splitting produces a finite signal at zero field with no such additional assumptions.

A second key advantage of CFs lies in the readout of magnetic information enabled by spin pumping.
In conventional antiferromagnets at zero field, degenerate magnon modes prevent the identification of the N\'{e}el configuration based on resonance measurements alone.
An external magnetic field lifts this degeneracy, but the resulting splitting is tied to the spin-up and spin-down directions rather than to the crystallographic sublattices. Therefore, when the N\'{e}el state is reversed, the roles of A and B are exchanged and a sublattice-equivalent antiferromagnet retains the same set of resonance frequencies.
In CFs, by contrast, the intrinsic sublattice asymmetry is pinned to the crystallographic sublattices through $K_{\rm A}\neq K_{\rm B}$ and produces N\'{e}el-state-dependent resonance frequencies.
Thus, the proposed readout relies on this N\'{e}el-state dependence, not merely on the lifting of a magnon degeneracy.
Since frequency-domain detection is intrinsically less sensitive to electrical noise than amplitude-based schemes, this mechanism offers a viable route to overcome the long-standing readout bottleneck in antiferromagnetic spintronics.
This distinction persists even under a finite magnetic field, provided that the field strength remains well below the spin-flip transition field.
We also note that the ``compensated ferrimagnet'' discussed in some prior works~\cite{Ohnuma2013} refers to a ferrimagnet at its magnetization-compensation temperature, which is a qualitatively different system from the intrinsically compensated ferrimagnet considered here.

\subsection{Experimental feasibility}

Finally, we comment on the experimental feasibility of realizing the sublattice asymmetry ($K_{\rm A}\neq K_{\rm B}$) central to our proposal.
Such asymmetry naturally arises in several material platforms, including Heusler alloys with symmetry-inequivalent magnetic sublattices~\cite{Xiao2025}, two-dimensional van der Waals magnets with broken out-of-plane mirror symmetry~\cite{Liu_PRL2025}, and chemically versatile organic antiferromagnets~\cite{Kawamura_PRL2024}.
Quantitatively, the average anisotropy parameter used in our calculations, $\bar{K}=5\times 10^{-3}J$, and a representative anisotropy difference $\Delta K=4\times 10^{-4}J$ ($\Delta K/\bar{K}=0.08$), correspond to $\bar{K}\sim 5$--$50\;\mu$eV/atom and $\Delta K\sim 0.4$--$4\;\mu$eV/atom for typical exchange couplings $J\sim 1$--$10$~meV. For comparison, the uniaxial anisotropy constant of hcp Co is $K_1 \approx 0.41$~MJ/m$^3$~\cite{Skomski2016}, corresponding to $\sim 28\;\mu$eV/atom (see Appendix~\ref{app:conversion} for the unit conversion). The average anisotropy $\bar{K}\sim 5$--$50\;\mu$eV/atom is comparable to or smaller than this value, and the required difference $\Delta K\sim 0.4$--$4\;\mu$eV/atom is smaller, confirming that our parameters are physically reasonable for 3$d$ magnetic compounds.
The largest value shown in Fig.~3, $\Delta K/\bar{K}=0.32$, corresponds to $\Delta K=1.6$--$16\;\mu {\rm eV}/{\rm atom}$ and is included to illustrate the scaling of the response, not as a minimum requirement. Since the spin current is approximately proportional to $\Delta K$, no threshold behavior at $\Delta K/\bar{K}=0.32$ is implied. Throughout the parameter range considered, $K_{\rm A}$ and $K_{\rm B}$ remain positive and much smaller than $J$, so the collinear ordered state remains stable.
In compounds containing 4$f$ rare-earth elements, uniaxial anisotropy energies can reach $\sim 1500\;\mu$eV/atom~\cite{Skomski2016}, more than an order of magnitude larger than that of 3$d$ compounds. 
Compensated ferrimagnets incorporating such elements may exhibit enhanced spin responses.

Regarding detectability, recent experiments on the longitudinal spin Seebeck effect (LSSE) combined with the inverse spin Hall effect (ISHE) have demonstrated nV-scale electrical sensitivity. For example, LSSE-based heat-flow sensors have achieved sensitivities of $V/q \approx 0.98~\mathrm{nV/(W/m^2)}$~\cite{Kirihara2016}, and spin Seebeck voltages of several tens of nV have been reported in ferrite systems~\cite{Ibarra2018}. 
Using the CF/FI current ratio of approximately \(1.5\times10^{-3}\) at \(k_{\rm B}T/J=0.5\) and \(\Delta K/\bar K=0.32\), a rough rescaling of the reported heat-flow sensitivity
gives voltages of approximately \(1.5\)–\(15~\mathrm{nV}\) for \(q\sim1\)–\(10~\mathrm{kW/m^2}\), assuming comparable thermal and interfacial conditions and spin-to-charge conversion efficiencies. This estimate is indicative and depends on the material and device geometry, but suggests that the signal may be experimentally accessible under favorable conditions.

\section{Summary}

We theoretically investigated the spin Seebeck effect and spin pumping in junctions between a compensated ferrimagnet and a normal metal.
We demonstrated that isotropic magnon spin splitting in compensated ferrimagnets generates a spin Seebeck signal that may be experimentally detectable under realistic experimental conditions, while retaining a small thermally induced net magnetization,
with its sign controlled by the two N\'{e}el configurations.
We further showed that spin pumping exhibits a N\'{e}el-state-dependent shift of the resonance frequency, enabling frequency-based readout of the two N\'{e}el states.
These findings establish compensated ferrimagnets with engineered sublattice asymmetry as suitable platforms for stray-field-free spintronic devices with reliable spin-current generation and state readout.

\begin{acknowledgments}
We thank Ryotaro Sano and Hisazumi Akai for valuable discussions.
This work was supported by the National Natural Science Foundation of China (NSFC) under Grant No. 12374126, by the Priority Program of Chinese Academy of Sciences under Grant No. XDB28000000, and by JSPS KAKENHI under Grants No. JP23H01839, No. JP23H03818, No. JP24K06951, No. JP24H00322, No. JP26H02144, and No. JP26K00652. T.M. was supported by JST FOREST under Grant No. JPMJFR236N. We acknowledge the use of the large language models ChatGPT (OpenAI, Astra) and Claude (Anthropic, Fable) in this work. The original numerical codes were written by the authors; the language models were used to check the consistency between the codes and the equations in the manuscript, to check the analytical derivations (which helped identify errors), to refactor and clean up the codes, to format the final output, and to prepare the documentation. All results and all modifications to the codes were verified by the authors, who take full responsibility for the content of this paper.
\end{acknowledgments}

\section*{DATA AVAILABILITY}

The data that support the findings of this study are openly available in Zenodo at Ref.~\cite{Lee2026Zenodo}.

\appendix

\section{Band Diagram}
\label{app:banddiagram}

\begin{figure}[b]
    \centering
    \includegraphics[width=0.8\linewidth]{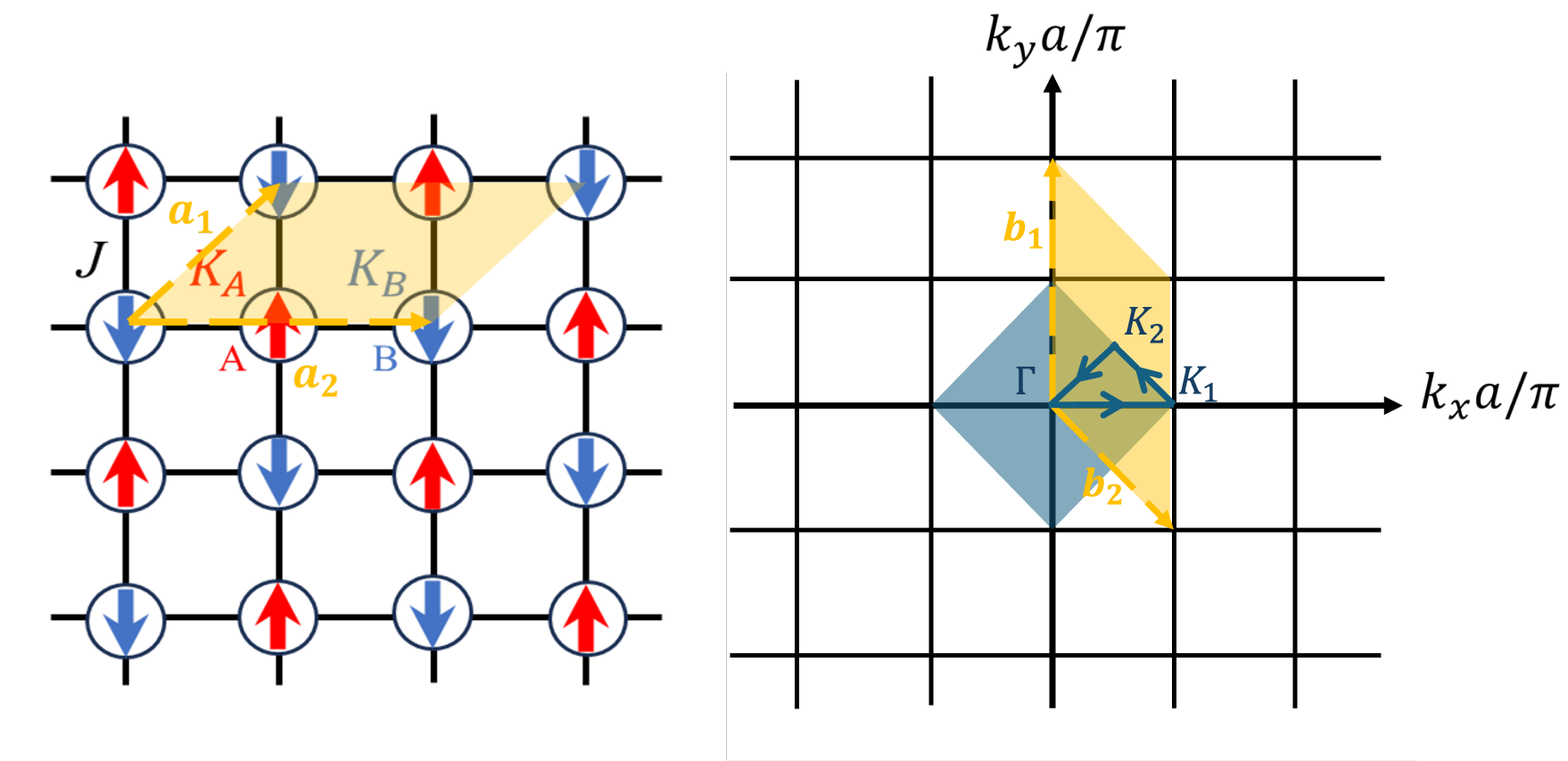}
    \caption{Real-space and reciprocal-space structure of the bipartite compensated ferrimagnet lattice. The first Brillouin zone is taken for a symmetric representation.}
    \label{fig:BZ}
\end{figure}

In this section, we briefly explain the band diagram momentum path taken in Fig.~\ref{fig:BandStructure} in the main text. We define the lattice vectors as $\bm a_1 = (a,a)$ and $\bm a_2 = (2a,0)$. The corresponding reciprocal vectors are $\bm b_1 = \pi/a(0,2)$ and $\bm b_2 = \pi/a(1,-1)$. The blue square represents the first Brillouin zone (see Fig.~\ref{fig:BZ}).

\section{Magnetization Dynamics of a Compensated Ferrimagnet}
\label{app:magdyn}

In this section, we provide a detailed analysis using the linearized Landau-Lifshitz-Gilbert (LLG) equation.
We consider a bipartite compensated ferrimagnet composed of two sublattices \(A\) and \(B\), described by the Hamiltonian
\begin{align}
H &= J \sum_{\langle ij \rangle} {\bm S}_i \cdot {\bm S}_j
-K_{\rm A} \sum_{i\in A} (S_i^z)^2
-K_{\rm B} \sum_{i\in B} (S_i^z)^2 \notag \\
&\quad - \sum_i {\bm S}_i \cdot {\bm B}_{\rm ac}(t),
\end{align}
where \(J\) ($>0$) is the antiferromagnetic exchange coupling and \(K_{{\rm A}}, K_{{\rm B}}\) ($>0$) are sublattice-dependent uniaxial anisotropies.
In this work, we only consider the case of $0<K_{\rm A},K_{\rm B} \ll J$.
The applied ac magnetic field is taken as
\begin{align}
\bm B_{\rm ac}(t)
= h\big(\cos\Omega t,-\sin\Omega t,0\big).
\end{align}
We note that the amplitude $h$ has a dimension of energy and is related to the amplitude of the external magnetic field $h_{\rm ac}$ as $h=-g\mu_{\rm B} h_{\rm ac}$, where $g$ ($>0$) is the Land\'{e} $g$ factor and $\mu_{\rm B}$ ($>0$) is the Bohr magneton.
Within a mean-field approximation, the system is reduced to an effective two-spin model, whose Hamiltonian is given by
\begin{equation}
H = zJ {\bm S}_{\rm A} \cdot {\bm S}_{\rm B}
-K_{\rm A} (S_{\rm A}^z)^2
-K_{\rm B} (S_{\rm B}^z)^2
- ({\bm S}_{\rm A} + {\bm S}_{\rm B})\cdot {\bm B}_{\rm ac}(t) ,
\end{equation}
where $z=4$ is the coordination number.
The magnetization dynamics obey the Landau-Lifshitz-Gilbert (LLG) equation
\begin{align}
\frac{d\bm S_\nu}{dt}
= \frac{1}{\hbar}\bm S_\nu\times\bm B_{{\rm eff},\nu}
-\frac{\alpha_{\rm G}}{S_0}
\bm S_\nu\times\frac{d\bm S_\nu}{dt}, \label{appeq:LLG}
\end{align}
where $\alpha_{\rm G}$ is the Gilbert damping constant, and the effective magnetic fields acting on the two sublattices, both of which have a dimension of energy, are given by
\begin{align}
{\bm B}_{\mathrm{eff},{\rm A}}(t) &= -zJ {\bm S}_{\rm B} + 2K_{\rm A} S_{\rm A}^z \hat{\bm z} + {\bm B}_{\rm ac}(t), \\
{\bm B}_{\mathrm{eff},{\rm B}}(t) &= -zJ {\bm S}_{\rm A} + 2K_{\rm B} S_{\rm B}^z \hat{\bm z}+ {\bm B}_{\rm ac}(t) .
\end{align}
The mean values of antiparallel equilibrium sublattice spins are denoted by ${\bm S}_{\rm A}^{(0)} = mS_0 \hat{\bm z}$ and ${\bm S}_{\rm B}^{(0)} = -mS_0 \hat{\bm z}$, where $m = \pm 1$ indicates the two possible spin configurations.
We consider small transverse fluctuations around a collinear equilibrium configuration,
\( {\bm S}_\nu= {\bm S}_\nu^{(0)}+\delta {\bm S}_\nu \)
($\nu = {\rm A},{\rm B}$) with \(|\delta S^{x,y}_\nu| \ll |S^{(0),z}_\nu|\).
Linearizing the LLG equation yields coupled equations for the transverse components.
To streamline the analysis, we proceed directly to circular components.
We introduce circular components for each sublattice,
\begin{equation}
\delta S_\nu^+ \equiv \delta S_\nu^x + i\,\delta S_\nu^y,
\qquad (\nu={\rm A}, {\rm B}).
\end{equation}
Then, the linearized transverse equations of the LLG equation can be written in closed form for $\delta S_\nu^+$ as:
\begin{align}
& \hbar \left(\begin{array}{cc} 1+im \alpha_{\rm G} & 0 \\
0 & 1-im\alpha_{\rm G} \end{array}\right) \left( \begin{array}{c}
\delta \dot{S}_{\rm A}^+ \\ \delta \dot{S}_{\rm B}^+ \end{array} \right) \notag \\
& = -i \left(\begin{array}{cc} \Omega_{\rm A} & \Lambda \\
-\Lambda & \Omega_{\rm B} \end{array}\right) \left( \begin{array}{c}
\delta S_{\rm A}^+ \\ \delta S_{\rm B}^+ \end{array} \right) + i h e^{-i\Omega t} \left( \begin{array}{c} mS_0 \\ -mS_0 \end{array} \right) ,
\end{align}
where
\begin{align}
& \Omega_{\rm A} = m S_0 (zJ + 2K_{\rm A} ), \quad
\Omega_{\rm B} = -m S_0 (zJ + 2K_{\rm B} ), \notag \\
& \Lambda = z J m S_0 .
\end{align}
We seek steady-state solutions of the form $\delta S_\nu^+(t) = \tilde S_\nu(\Omega)\,e^{-i\Omega t}$ ($\nu={\rm A}, {\rm B}$).
Substituting into the equations of motion yields a \(2\times2\) algebraic system,
\begin{widetext}
\begin{align}
\left( \begin{array}{cc}
-\hbar \Omega(1+i m\alpha_{\rm G} )+\Omega_{\rm A} &
  +\Lambda \\
  -\Lambda & -\hbar \Omega(1-im \alpha_{\rm G})+\Omega_{\rm B} \end{array} \right) \left( \begin{array}{c} \tilde{S}_{\rm A}(\Omega) \\ \tilde{S}_{\rm B}(\Omega) \end{array}\right) &= h \left( \begin{array}{c} mS_0 \\ -m S_0 \end{array} \right).
\end{align}
\end{widetext}
Introducing $D_{\rm A}(\Omega) =-\hbar
\Omega(1+im \alpha_{\rm G} )+\Omega_{\rm A}$ and $D_{\rm B}(\Omega) =
-\hbar \Omega(1-im \alpha_{\rm G} )+\Omega_{\rm B}$,
the amplitudes of the spin precession on the A and B sublattices are given by
\begin{align}\label{eq:inplaneSASB2}
\left(\begin{array}{c}
\tilde{S}_{\rm A}(\Omega) \\ \tilde{S}_{\rm B}(\Omega) \end{array}\right)
= \frac{h}{\Delta(\Omega)}
\left( \begin{array}{cc} D_{\rm B}(\Omega) & -\Lambda \\ \Lambda & D_{\rm A}(\Omega) \end{array} \right) \left( \begin{array}{c} mS_0 \\ -mS_0 \end{array} \right),
\end{align}
where $\Delta(\Omega) = D_{\rm A}(\Omega)D_{\rm B}(\Omega)+\Lambda^2$ denotes the determinant.
The magnon resonance frequencies are determined by \( \Delta(\Omega)=0 \) and $\alpha_{\rm G}= 0$, which yield
\begin{align}
\Delta(\Omega)
&= (\hbar \Omega)^2
- 2mS_0(K_{\rm A}-K_{\rm B})  \hbar \Omega \notag \\
&\quad - S_0^2 \{2zJ (K_{\rm A} + K_{\rm B})
+ 4K_{\rm A} K_{\rm B}\}= 0, \\
\hbar \Omega
&=  \hbar \omega_{\pm} \notag \\
&= mS_0 (K_{\rm A}-K_{\rm B}) \notag \\
&\quad \pm S_0 \sqrt{(K_{\rm A}+K_{\rm B})^2
+ 2zJ(K_{\rm A}+K_{\rm B})} .
\end{align}
Since $K_{\rm A},K_{\rm B}\ll J$, the square-root term dominates over the linear term
$\propto (K_{\rm A}-K_{\rm B})$, and thus we obtain $\omega_-<0<\omega_+$.

\begin{figure}[b]
\centering
\includegraphics[width=\linewidth]{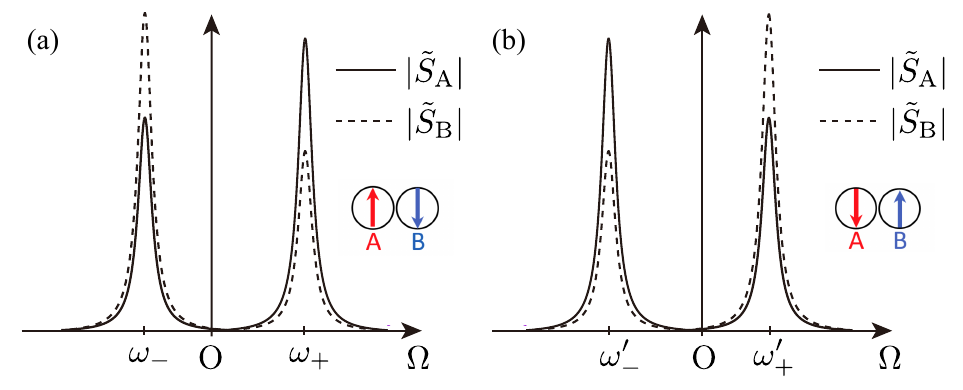}
\caption{Schematic illustration of the spin precession amplitudes $|\tilde{S}_{\rm A}(\Omega)|$ (solid lines) and $|\tilde{S}_{\rm B}(\Omega)|$ (dashed lines) for $K_{\rm A}>K_{\rm B}$. (a) The $m=+1$ configuration. (b) The $m=-1$ configuration. The resonance frequencies satisfy $|\omega_-|<|\omega_+|$ and $|\omega_-'|>|\omega_+'|$. The resonance positions are schematic and not drawn to scale. The curves illustrate the qualitative ordering of the peak heights, rather than their quantitative ratios.}
\label{fig:spinprecession}
\end{figure}

Using Eq.~\eqref{eq:inplaneSASB2}, we get
\begin{align}
    \tilde{S}_{\rm A}(\Omega) &= \frac{h}{\Delta(\Omega)}mS_0(D_{\rm B}(\Omega) + \Lambda), \\
    \tilde{S}_{\rm B}(\Omega) &= \frac{h}{\Delta(\Omega)}mS_0(\Lambda-D_{\rm A}(\Omega)).
\end{align}
The relative magnitudes of $|\tilde{S}_{\rm A}(\Omega)|$ and $|\tilde{S}_{\rm B}(\Omega)|$ can be evaluated using
\begin{align}
    D_{\rm B}(\Omega) + \Lambda &= -\hbar \Omega - 2mS_0 K_{\rm B} + i \hbar \Omega m\alpha_{\mathrm G}, \\
    D_{\rm A}(\Omega) - \Lambda &= - \hbar \Omega + 2mS_0 K_{\mathrm A} - i \hbar \Omega m\alpha_{\rm G}.
\end{align}
When $m=1$, corresponding to the configuration in which the spins on sublattice A (B) align along the $+z$ ($-z$) direction, one finds $|\tilde{S}_{\rm A}(\omega_+)| > |\tilde{S}_{\rm B}(\omega_+)|$ and $|\tilde{S}_{\rm A}(\omega_-)| <  |\tilde{S}_{\rm B}(\omega_-)|$.
These inequalities correspond to different amplitudes of the precession shown in Fig.~2(b1) in the main text.
On the other hand, when $m=-1$, corresponding to the configuration in which the spins on sublattice A (B) align along the $-z$ ($+z$) direction, one finds $|\tilde{S}_{\rm A}(\omega_+)| < |\tilde{S}_B(\omega_+)|$ and $|\tilde{S}_{\rm A}(\omega_-)| > |\tilde{S}_B(\omega_-)|$, which correspond to the precession amplitudes shown in Fig.~2(b2) in the main text.
These features are illustrated in Fig.~\ref{fig:spinprecession}.

Finally, we examine the phase relation in the spin precession between sublattices A and B, assuming $K_{\rm A}, K_{\rm B} \ll J$ and $\alpha_{\rm G} \ll 1$, under which the eigenfrequencies are given by $\hbar \omega_\pm \simeq \pm S_0 \sqrt{2zJ(K_{\rm A}+K_{\rm B})}$.
Therefore, we obtain $\tilde{S}_{\rm A}(\omega_+)/\tilde{S}_{\rm B}(\omega_+) \simeq \tilde{S}_{\rm A}(\omega_-)/\tilde{S}_{\rm B}(\omega_-) \simeq -1$.
This result indicates the anti-phase relation in the precession of the spins between the sublattices A and B, as shown in Fig.~2(b1) and Fig.~2(b2) in the main text.

\section{Green's Function}\label{appex:GreenFunction}

In this appendix, we summarize the definitions of the Green's functions used throughout the paper for consistency.
We define the Green's function for $\nu,\mu = A,B$ as
\begin{align}\label{eq:Green'sfunction}
    G_{\nu\mu}(\bm{k},\tau,\tau') \equiv -\frac{i}{\hbar} \bigg \langle \mathcal{T}_K S^+_{\nu \bm k}(\tau)S^-_{\mu \bm k}(\tau')\bigg \rangle.
\end{align}
where $\tau,\tau'$ denote time on the Keldysh contour and $\mathcal{T}_K$ is the contour-ordering operator.
We also define the spin susceptibility as
\begin{align}
    \chi(\bm{k},\tau,\tau') &\equiv \frac{i}{\hbar N_{\rm NM}} \bigg \langle \mathcal{T}_K s^+_{\bm k}(\tau) s^-_{\bm k}(\tau')\bigg \rangle,
\end{align}
where $s^\pm_{\bm k}$ are the spin ladder operators of the normal metal.
For a detailed calculation, see Refs.~\cite{Ominato2025,kato19}.
Note that the definitions of the lesser Green's functions for the normal metal and the compensated ferrimagnet differ by a sign convention.
With the present convention, the positive- and negative-frequency magnon poles carry positive and negative spectral weights, respectively.
We can consistently derive the explicit form of the retarded Green's function for the sublattice model as
\begin{align}
    & G_{\rm AA}^R(\bm k,\omega)/2S_0 \notag \\
    & \quad = \frac{|u_{\bm k}|^2}{\hbar \omega - \hbar \omega_{+,\bm{k}} + i0^+} - \frac{|v_{\bm k}|^2}{\hbar \omega - \hbar \omega_{-,\bm k} + i0^+} , \\
    & G_{\rm BB}^R(\bm k,\omega)/2S_0 \notag \\
    &\quad = \frac{|v_{\bm k}|^2}{\hbar \omega - \hbar \omega_{+,\bm{k}} + i0^+} - \frac{|u_{\bm k}|^2}{\hbar \omega - \hbar \omega_{-,\bm k} + i0^+}\\
    &G_{\rm AB}^R(\bm k,\omega)/2S_0=G_{\rm BA}^R(\bm k,\omega)/2S_0 \notag \\
    &\quad = -\frac{u_{\bm k}v_{\bm k}}
    {\hbar\omega-\hbar\omega_{+,\bm k}+i0^+}
     +\frac{u_{\bm k}v_{\bm k}}
    {\hbar\omega-\hbar\omega_{-,\bm k}+i0^+}.
\end{align}
To account for weak damping in spin pumping experiments, we broaden each magnon pole by replacing a positive infinitesimal with a positive linewidth \(\hbar \Gamma_{\pm,\bm k}\). 
Here, \(\Gamma_{\pm,\bm k}>0\) is the magnon decay rate, and \(\hbar\Gamma_{\pm,\bm k}\) is the corresponding energy half-width.
This prescription places both poles in the lower half of the complex-frequency plane and preserves the retarded character of the Green's functions. We retain the undamped pole residues, which is appropriate near well-resolved resonances in the weak-damping limit. Using the LLG equation introduced in Appendix~\ref{app:magdyn}, the linewidth \(\Gamma_{\pm,\bm k}\) for the uniform spin precession (${\bm k}={\bm 0}$) can be related to the Gilbert damping constant $\alpha_{\rm G}$ as
\begin{align}
\Gamma_{\pm,\bm0}
&=\tilde\alpha_{\rm G}|\omega_{\pm,\bm0}| + {\cal O}(\alpha_{\rm G}^2), \\
\tilde\alpha_{\rm G}
&=\alpha_{\rm G}
\frac{A+B}{\sqrt{(A+B)^2-4C_{\bm0}^{\,2}}} ,
\end{align}
where $A$, $B$, and $C_{\bm k}$ are defined in Sec.~\ref{subsec:CFerri}.
Here, $\tilde{\alpha}_{\rm G}$ is the dimensionless effective damping constant, which is proportional to the Gilbert damping constant $\alpha_{\rm G}$ to leading order in the Gilbert damping.

We define the lesser Green's function as
\begin{align}
\label{Glesser}
G^{<}_{\nu\nu}(\bm{k},t) &= -\frac{i}{\hbar}\langle S_{\nu \bm{k}}^{-}(0) S_{\nu \bm{k}}^+(t) \rangle ,\\
\label{chilesser}
\chi^{<}(\bm{q},t) &= \frac{i}{\hbar N_{\rm NM}} \langle s_{\bm{q}}^{-}(0)s_{\bm{q}}^+(t) \rangle .
\end{align}
When the compensated ferrimagnet and normal metal are decoupled and in thermal equilibrium at the temperatures $T_{\rm CF}$ and $T_{\rm NM}$, the lesser component can be related to the retarded component through the fluctuation-dissipation theorem as~\cite{Stefanucci2013}
\begin{align}
G^{<}_{\nu\nu}(\bm{k},\omega) &=2i f(\hbar \omega,T_{\rm CF}) \text{Im}\, [G^{R}_{\nu\nu}(\bm{k},\omega)],\label{eq:generalflucdissGF} \\
\chi^{<}(\bm{q},\omega) &= 2if(\hbar \omega,T_{\rm NM}) \text{Im}\, [\chi^{R}(\bm{q},\omega)],
\label{eq:generalflucdisssuscep}
\end{align}
where $f(\varepsilon,T)=(e^{\varepsilon/k_{\rm B}T}-1)^{-1}$ denotes the Bose distribution function.

Next, we evaluate the microwave-induced correction to the lesser Green's function using the broadened retarded Green's functions introduced above.
The lesser Green's function for spin pumping is obtained from the second-order perturbation expansion in the external field $V(t)$ and is given explicitly by
\begin{align}\label{eq:pumpinglesserGF}
    \delta G^<_{\nu\nu}(\bm{k},\omega) &= -\frac{i \pi \hbar \gamma^2 h_{ac}^2 N_{\rm CF}}{2}\,\delta_{\bm k,\bm 0}\,\delta(\omega - \Omega) \notag \\
    &\quad \times \left|\sum_{\mu}G^R_{\nu\mu}(\bm k,\omega)\right|^2 , \quad (\nu = {\rm A}, {\rm B}), 
\end{align}
A global spin flip by $180^\circ$ leads to a change in the resulting Green's function.
The Holstein-Primakoff transformation changes to
\begin{align}
S_{\nu \bm{R}}^z &\approx
\begin{cases}
-S_0 + a_{\bm{R}}^{\dagger}a_{\bm{R}} & \nu = \rm A , \\
S_0 - b_{\bm{R}}^{\dagger}b_{\bm{R}} & \nu = \rm B ,
\end{cases} \label{eq:spinwaveapprox1/1} \\
S_{\nu \bm{R}}^+ & \approx
\begin{cases}
\sqrt{2S_0}a_{\bm{R}}^\dagger & \nu = \rm A , \\
\sqrt{2S_0}b_{\bm{R}} & \nu = \rm B ,
\end{cases}
\label{eq:spinwaveapprox2/1} \\
S_{\nu \bm{R}}^- & = (S_{\nu \bm{R}}^+)^\dagger ,
\label{eq:spinwaveapprox3/1}
\end{align}
By applying the definition of the Green's function in Eq.~(\ref{eq:Green'sfunction}), we obtain the retarded Green's functions as
\begin{align}
    &G_{\rm AA}^R(\bm k,\omega) /2S_0\notag \\
    &\quad = \frac{|v_{\bm k}|^2}{\hbar \omega - \hbar \omega_{+,\bm{k}}' + i \hbar\Gamma_{+,\bm k}'} - \frac{|u_{\bm k}|^2}{\hbar \omega - \hbar \omega_{-,\bm k}' + i \hbar\Gamma_{-,\bm k}'}, \\
    & G_{\rm BB}^R(\bm k,\omega) /2S_0\notag \\
    &\quad = \frac{|u_{\bm k}|^2}{\hbar \omega - \hbar \omega_{+,\bm{k}}^{\prime} + i \hbar\Gamma_{+,\bm k}'}- \frac{|v_{\bm k}|^2}{\hbar \omega - \hbar \omega_{-,\bm k}' + i \hbar\Gamma_{-,\bm k}'},\\
    &G_{AB}^{R}(\bm k,\omega)/2S_0=G_{BA}^{R}(\bm k,\omega)/2S_0 \notag \\
&\quad = -\frac{u_{\bm k}v_{\bm k}}
{\hbar\omega-\hbar\omega_{+,\bm k}'
 +i\hbar\Gamma_{+,\bm k}'}+\frac{u_{\bm k}v_{\bm k}}
{\hbar\omega-\hbar\omega_{-,\bm k}'
 +i\hbar\Gamma_{-,\bm k}'}.
\end{align}
where $\omega_{\pm,{\bm k}}' =-\omega_{\mp,{\bm k}}$ and $\Gamma_{\pm,{\bm k}}'=\Gamma_{\mp,{\bm k}}$ follow from the relationship between the two spin configurations.

\section{Detailed Derivation of the Spin Current}\label{appex:derivationofSpinCurrent}

Here, we derive the general expression of the spin current.
The spin current is calculated within the Keldysh formalism as~\cite{Bruus,kato19}
\begin{widetext}
\begin{align}\label{eq:spincurrentstep1}
\langle I_S \rangle
&= -\text{Im}\,\bigg[ \sum_{\nu \bm{R}}J_{\nu {\bm R},{\bm r}} \bigg\langle S^+_{\nu \bm{R}}(\tau_1)s_{\bm r}^-(\tau_2) \exp\bigg(-\frac{i}{\hbar}\int_C d\tau H_{\text{int}}(\tau) \bigg)\bigg\rangle_0\bigg] \notag \\
&= \frac{1}{2\hbar} \, \text{Re}\,\bigg[ \sum_{\nu \bm{R}} \sum_{\nu^{\prime} \bm{R}^{\prime}} J_{\nu {\bm R},{\bm r}} J_{\nu^{\prime} {\bm R}^{\prime}, {\bm r}'}^{\ast} \int_C d\tau \, \biggl\langle \mathcal{T}_K S_{\nu \bm{R}}^+(\tau_1) s_{{\bm r}}^-(\tau_2) S_{\nu^{\prime} \bm{R}^{\prime}}^-(\tau) s_{{\bm r}'}^+(\tau)\biggr\rangle_0\bigg] ,
\end{align}
\end{widetext}
where $\langle \cdots \rangle_0$ represents the thermal average with respect to $H_0$, $\tau_1=(t,+)$ and $\tau_2=(t,-)$ denote contour times on the Keldysh contour, $\int_C d\tau$ indicates integration along the contour, and $A(\tau)=e^{iH_0 \tau/\hbar} A e^{-iH_0 \tau/\hbar}$ denotes the real-time evolution under the Hamiltonian $H_0=H_{\rm CF}+H_{\rm NM}$.
Using the disorder average $\langle J_{\nu{\bm R}} J_{\nu^{\prime}{\bm R}'}^{\ast}\rangle_{\rm imp} = |J_{\rm CF}|^2 \delta_{\bm{R},\bm{R}^{\prime}} \delta_{\nu,\nu^{\prime}}$, we obtain
\begin{widetext}
\begin{align}\label{eq:spincurrentstep2}
\langle I_S\rangle &= \frac{\hbar}{2} \,\text{Re}\,\bigg[ \sum_{\nu \bm{R}} \sum_{\bm{k} ,\bm{q}} \frac{|J_{\rm CF}|^2}{N_{\rm CF}N_{\rm NM}} \int_C d\tau \, G_{\nu \nu} (\bm{q},\tau_1,\tau) \chi(\bm{k},\tau,\tau_2)\bigg] \notag \\
&= \frac{\hbar}{2}\sum_{\bm{k} ,\bm{q}}  \,\text{Re}\, \bigg[  \sum_{\nu \bm{R}}  \frac{|J_{\rm CF}|^2}{N_{\rm CF}N_{\rm NM}}\int_{-\infty}^{\infty} \frac{d\omega}{2\pi} \, \bigg(\chi^{<}(\bm{k},\omega)G_{\nu \nu}^{R}(\bm{q},\omega) + \chi^{A}(\bm{k},\omega) G_{\nu \nu}^{<}(\bm{q},\omega)\bigg)\bigg],
\end{align}
\end{widetext}
where $G_{\nu\nu} (\bm{q},\omega)$ and $\chi(\bm{k},\omega)$ are the nonequilibrium Green's functions for the compensated ferrimagnet and normal metal, respectively, and the Langreth rule~\cite{Stefanucci2013} has been used in the second equality.
Finally, substituting the fluctuation-dissipation relations in Eqs.~\eqref{eq:generalflucdissGF}and \eqref{eq:generalflucdisssuscep},we obtain the general expression for the spin current,
\begin{widetext}
\begin{align}
\label{eq:generalspincurrent}
    \langle I_S \rangle = \frac{\hbar |J_{\rm CF}|^2}{N_{\rm NM}N_{\rm CF}} \sum_{\bm k,\bm q} \sum_{\nu} N_{{\rm int},\nu} \int_{-\infty}^{\infty} \frac{d\omega}{2 \pi} \, \text{Im} \, G_{\nu\nu}^R(\bm k,\omega)  \, \text{Im}\,\chi^R(\bm q,\omega)\bigg(f_{\nu}^{\rm CF}(\hbar \omega) - f^{\rm NM}(\hbar \omega)\bigg),
\end{align}
\end{widetext}
where $N_{{\rm int},\nu}$ is the number of the interfacial bonds for the sublattice $\nu$.
For simplicity, we assume that the number of the interfacial bonds are the same for the two sublattices, i.e., $N_{{\rm int},\nu}=N_{\rm int}/2$, where $N_{\rm int}$ is the total number of the interfacial bonds.
The local spin susceptibility is calculated in the wide-band limit as
\begin{align}
\frac{1}{N_{\rm NM}} \sum_{\bm q} {\rm Im} \, \chi^R({\bm q},\omega) = \pi N(0)^2 \hbar \omega,
\end{align}
where $N(0)$ is the density of states per spin and per unit cell for the normal metal at the Fermi energy.
Substituting this expression into Eq.~\eqref{eq:generalspincurrent}, we finally obtain Eq.~\eqref{eq:generalspincurrent_main}.

\section{Magnetization}
\label{app:magnetization}

\begin{figure}[b]
\centering
\includegraphics[width=0.8\linewidth]{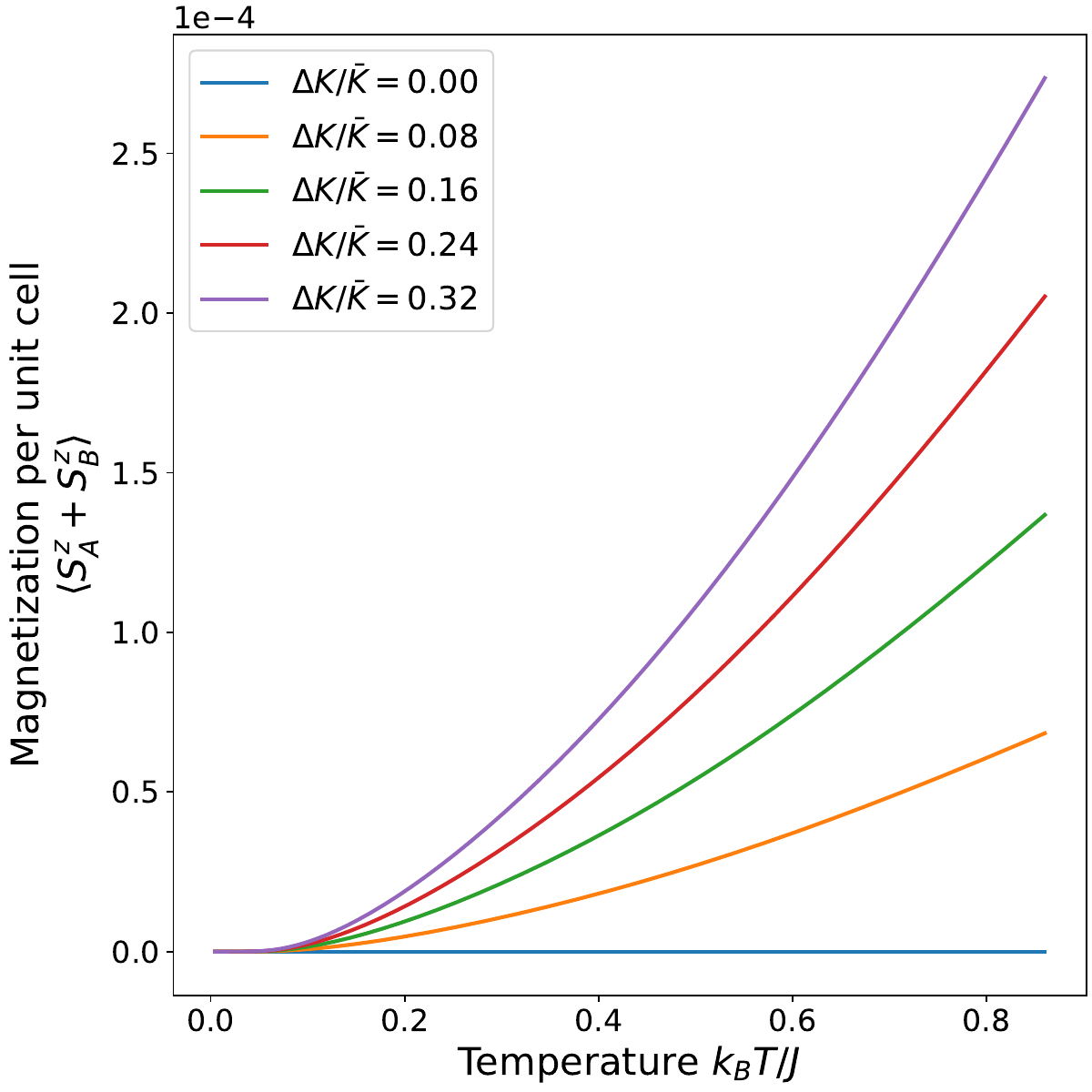}
\caption{Temperature dependence of the magnetization per unit cell, $\langle S^z_{\rm A}+S^z_{\rm B}\rangle$, for the compensated ferrimagnet.
The parameters are chosen as $K_{\rm A}=\bar{K}+\Delta K/2$ and $K_{\rm B}=\bar{K}-\Delta K/2$, and the solid lines correspond to $\Delta K/\bar{K} = 0, 0.08, 0.16, 0.24$, and 0.32 for $\bar{K} = 0.005J$ and $S_0=1$.}
\label{fig:magnetization}
\end{figure}

As discussed in Appendix~\ref{app:magdyn}, the imbalance of the precession amplitudes on the two sublattices implies that the two magnon branches carry opposite $z$ polarizations, i.e., opposite contributions to $S^z_{\rm A}+S^z_{\rm B}$.
As a result, the energy splitting between the two branches yields a finite net magnetization per unit cell at finite temperature.
Using $S^z_{\rm A}+S^z_{\rm B}=N_{\rm CF}^{-1} \sum_{\bm k}(-a_{\bm k}^\dagger a_{\bm k}+b_{\bm k}^\dagger b_{\bm k})$, we obtain
\begin{align}
\langle S^z_{\rm A} + S^z_{\rm B} \rangle
= \frac{1}{N_{\rm CF}}\sum_{\bm k}\Bigl[ f(-\hbar \omega_{-,\bm k}) - f(\hbar \omega_{+,\bm k}) \Bigr],
\end{align}
where $f(x)=1/(e^{\beta x}-1)$ is the Bose distribution function and we used the convention $\omega_{-,\bm k}<0<\omega_{+,\bm k}$.
The temperature dependence of the magnetization per unit cell $\langle S^z_{\rm A} + S^z_{\rm B} \rangle$ is shown in Fig.~\ref{fig:magnetization}.
This expression should be understood as a low-temperature
linear-spin-wave estimate for the compensated ferrimagnet considered
in this work, where $\Delta K\neq0$. When $\Delta K=0$, the model
reduces to a sublattice-equivalent antiferromagnet. In that limit,
the A$\leftrightarrow$B symmetry forbids any net magnetization, so
$\langle S^z_{\rm A}+S^z_{\rm B}\rangle=0$ is the symmetry-required
result, including at infinite temperature.
The infinite-temperature limit itself is outside the validity of
linear spin-wave theory, which assumes a magnetically ordered state
and a small magnon density. Critical fluctuations near the magnetic
transition temperature are also neglected.

\section{Conversion of anisotropy constants to $\mu$eV/atom}
\label{app:conversion}

We convert the anisotropy constant $K_1$ in SI units (MJ/m$^3$) to energy per atom using
\begin{align}
E\;[\mu\text{eV/atom}]
= \frac{K_1\;[\text{MJ/m}^3]\times\Omega\;[\text{\AA}^3/\text{atom}]}{0.1602},
\end{align}
where $\Omega$ is the volume per atom and the factor $0.1602$ arises from $1\;\text{eV}=1.602\times10^{-19}\;\text{J}$.
For hcp Co, the lattice constants are $a=2.51\;\text{\AA}$ and $c=4.07\;\text{\AA}$~\cite{Kittel2005}. The unit-cell volume is $V_{\rm cell}=(\sqrt{3}/2)\,a^2 c\sim 22.2\;\text{\AA}^3$. Since the hcp unit cell contains two Co atoms, $\Omega=V_{\rm cell}/2\sim 11.1\;\text{\AA}^3/\text{atom}$.
With $K_1\sim 0.41\;\text{MJ/m}^3$~\cite{Skomski2016}, we obtain $K_1\sim 28\;\mu\text{eV/atom}$.

As an example of a 4$f$ rare-earth system, we consider SmCo$_5$. SmCo$_5$ has the CaCu$_5$-type structure (space group $P6/mmm$) with lattice constants $a=4.99\;\text{\AA}$ and $c=3.98\;\text{\AA}$~\cite{Coey2010}.
The unit cell contains one formula unit (1 Sm + 5 Co = 6 atoms). The unit-cell volume is $V_{\rm cell}=(\sqrt{3}/2)\,a^2 c\sim 85.7\;\text{\AA}^3$ and $\Omega=V_{\rm cell}/6\sim 14.3\;\text{\AA}^3/\text{atom}$.
With $K_1\sim 17\;\text{MJ/m}^3$ at room temperature~\cite{Skomski2016}, we obtain $K_1\sim 1500\;\mu\text{eV/atom}$.

\bibliography{reference}

\end{document}